\documentclass[acmlarge]{acmart}
\makeatletter
\newcommand{\myconfshort}{\acmConference@shortname}
\newcommand{\myconffull}{\acmConference@name}
\newcommand{\myconfdate}{\acmConference@date}
\newcommand{\myconfloc}{\acmConference@venue}
\AtBeginDocument{
  \fancypagestyle{firstpagestyle}{
    \fancyhead{}%
    \fancyfoot[C]{}%
  }
  \fancyhf{}
  \fancyhead[LO]{\@headfootfont\shorttitle}%
  \fancyhead[RE]{\@headfootfont\@shortauthors}%
  \fancyhead[LE]{\@headfootfont\footnotesize \myconfshort, \myconfdate, \myconfloc}%
  \fancyhead[RO]{\@headfootfont\footnotesize \myconfshort, \myconfdate, \myconfloc}%
  \fancyfoot[C]{}%
}
\makeatother
\acmBooktitle{\conffull\@ (\confshort), \confdate, \confloc}
\acmBooktitle{\conffull\@ (\confshort), \confdate, \confloc}

\usepackage{enumitem}
\usepackage{tabularx}
\usepackage{enumitem}
\usepackage{booktabs}
\usepackage{xurl}
\usepackage{ccicons}

\copyrightyear{2026}
\acmYear{2026}
\setcopyright{cc}
\setcctype{by}
\acmConference[FAccT '26]{The 2026 ACM Conference on Fairness, Accountability, and Transparency}{June 25--28, 2026}{Montreal, QC, Canada}
\acmBooktitle{The 2026 ACM Conference on Fairness, Accountability, and Transparency (FAccT '26), June 25--28, 2026, Montreal, QC, Canada}
\acmDOI{10.1145/3805689.3806744}
\acmISBN{979-8-4007-2596-8/2026/06}

\begin{document}

\title[Cripping AI]{Cripping AI: Reimagining AI Through Lived Disability Experiences}
%Cripping AI: A Justice-Oriented Framework Toward Disability-Centered AI [this mentions both disability justice and disability centered..]
%%Cripping AI: A Theorectical Framework Toward Disability-Centered AI

\author{Xinru Tang}
\orcid{0000-0001-6426-1363}
\affiliation{\institution{University of California, Irvine}\city{Irvine}\state{California}\country{USA}}
\affiliation{\institution{AImpower.org}\city{Mountain View}\state{California}\country{USA}}
\email{xinrut1@uci.edu}

\author{Ting-an Lin}
\orcid{0000-0002-3519-6221}
\affiliation{\institution{University of Connecticut}\city{ Storrs}\state{Connecticut}\country{USA}}
\email{tinganlin@uconn.edu}

\author{Jingjin Li}
\orcid{0000-0003-0193-7228}
\affiliation{\institution{AImpower.org}\city{Mountain View}\state{California}\country{USA}}
\email{jingjin@aimpower.org}

\author{Shaomei Wu}
%\authornote{Corresponding author}
\orcid{0000-0003-1104-4116
}
\affiliation{\institution{AImpower.org}\city{Mountain View}\state{California}\country{USA}}
\email{shaomei@aimpower.org}
\renewcommand{\shortauthors}{Tang et al.}

%New for this year, we are requiring all authors to include a disclosure of generative AI use in the Endmatter, even if you did not use generative AI. See the Author Guide for more details.

\begin{abstract}
Drawing on crip theory, this paper proposes \textit{cripping AI} as a guiding framework to center lived disability experiences in AI research and development. Moving beyond calls to make AI ``accessible'' to people with disabilities, cripping AI seeks to: (1) reveal and dismantle ableist assumptions embedded in how AI is imagined, designed, and evaluated; (2) center disabled ways of knowing (i.e., cripistemologies); (3) respect disabled labor in co-creating accessible practices. We demonstrate how to apply our framework with three cases: deafness and sign language AI, blindness and visual assistive AI, and stuttering and speech AI. We end by outlining three directions for future work, including cripping AI with diverse human bodyminds, across the entire AI pipeline and ecosystem, and in collaboration with other justice-oriented AI efforts.
%how to advance the goal of cripping AI in dialogue with diverse human bodyminds, throughout the entire AI lifecycle and ecosystem, and along with other critical discourses.
\end{abstract}




\maketitle

\section{Introduction}
\begin{quote}
    Today, billions of dollars are poured into projects seeking to increase inclusion. It's not that I don't appreciate it when a restaurant has a Braille menu...But the way those things are lobbied for, funded, designed, implemented, and used revolves around the assumption that there's only one world... --\textit{Against Access}, John Lee Clark~\cite{clark-against-access}
\end{quote} 
Decades of disability activism have established accessibility and disability inclusion as core principles of technology development in both Web standards~\cite{web-accessibility} and laws~\cite{eu-a11y, china-laws-a11y, ada}. These influences have now positioned disability as a vital dimension of responsible AI~\cite{guo2020toward, whittaker2019disability, trewin2019considerations, google-signgemma, google-relate, apple-new-features, blind-meta, nvidia-sign}. According to the World Health Organization (WHO), around 1.3 billion people worldwide experience significant disability, representing approximately 16\% of the global population~\cite{who-disability}. People with disabilities (PWDs)\footnote{We use both person-first (people with disabilities) and identity-first (disabled people) interchangeably to recognize diverse preferences of language in disabled communities.}, therefore, constitute perhaps the largest and most diverse minority group. Aging, illness, and injury mean that anyone can experience disability at any point in their lives. Nonetheless, disability continues to be systematically marginalized in AI development. Growing evidence shows that AI models exhibit both performance degradation and bias in disability-related tasks~\cite{guo2020toward, whittaker2019disability, trewin2019considerations}. For example, foundation models trained on large-scale data often reproduce stereotypes about disabled people~\cite{mack2024representation, gadiraju2023disability, park2025paradox, glazko2024resume} and fail on disability-specific tasks, such as recognizing white canes in images~\cite{massiceti2024explaining} or transcribing stuttered speech~\cite{sridhar2025jjj, li2024towards}. In response, extensive efforts have been made to improve AI for PWDs, ranging from dataset curation~\cite{li2024towards, li2025collective, gong202470, desai2023asl, theodorou2021disability, massiceti2021orbit, sharma2023disability, garg2025s} to system building~\cite{huh2023genassist, chen2025surfacing, theodorou2021disability, lee-altcanvas, zhang2025towards}.

While these initiatives represent important progress toward making AI more accessible, there remains a lack of clear guidance on how to make AI truly supportive of disability. Scholars and activists have raised concerns that many accessibility efforts continue to assume and reinforce an ableist worldview. For example, many assistive technologies are designed around the assumption that able-bodiedness is the `ideal', whereas disability is a medical deficit, or a problem that needs to be fixed through these technologies~\cite{shew2023against, mankoff2010disability, spiel2019agency, spiel2022adhd}. As a result, these technologies might still reinforce the ableist hierarchy and treat PWDs as passive objects of protection. These reflections thus lead to Deafblind activist John Lee Clark's provocative call to ``against access,''~\cite{clark-against-access} and autistic scholar J. Logan Smilges's claim that ``if anything, I wanted less access.''~\cite{smilges2023crip}

To truly center disability in AI development, this paper takes one step further from \textit{making AI accessible} toward \textit{cripping AI} as a guiding framework. We turn to \textit{crip theory} as it is a key intellectual tradition that emphasizes dismantling ableist assumptions and centering disabled people's lived experiences~\cite{kafer2013feminist, mcruer2008crip}. We propose \textit{cripping AI} with three interconnected core tenets:
\setlist{nosep}
\begin{enumerate}
    \item \textbf{Expose disability politics:} Rather than upholding a rigid classification of `disabled' or `non-disabled,' research should examine how AI practices construct certain human bodyminds as `disabled' and develop strategies to affirm disabled experiences. We use the term ``bodymind'' following the tradition of critical disability studies to acknowledge that human bodymind acts as one~\cite{price2015bodymind}.
    \item \textbf{Honor cripstemologies:} Instead of building systems that treat able-bodied perceptions or knowledge as the `ground truth,' development should center the knowledge developed through disabled bodyminds or from disabled communities.
    \item \textbf{Respect crip labor}: Instead of treating disabled people as passive recipients of protection, research should recognize that accessibility is an ongoing, friction-filled, collective work in which disabled people play a key role. For example, attention should be given to how AI systems redistribute the labor involved in co-creating access.
\end{enumerate}
The rest of the paper proceeds as follows. Section \ref{sec::related-work} situates cripping AI within the related literature, including efforts to make AI accessible and crip HCI. Section \ref{sec::cripping} introduces the three core tenets of cripping AI. Section \ref{sec::cases} presents three case studies drawing from our own research to illustrate how to effectively put cripping AI in practice. Finally, Section \ref{sec::discussion} explores future directions to advance our framework.

\vspace{-1em}
\section{Related Work and Background}
\label{sec::related-work}
\subsection{Limitations of `Making AI Accessible'}
The wide recognition of disability exclusion in AI development has spurred increasing efforts to make AI accessible, i.e., to design and develop AI systems that PWDs can use or that meet their needs. The primary efforts include curating more representative datasets with affected communities~\cite{desai2023asl, bragg2022crowdsourcing, gong202470, gurari2018vizwiz, gurari2020captioning}, involving disabled people in co-design iterations and model evaluations~\cite{huh2023genassist, chang2024world, bennett2025toward, gadiraju2023disability, mack2024representation}, and assessing model performance and downstream impacts on PWDs~\cite{sridhar2025jjj, massiceti2024explaining, phutane2025cold, mack2024representation, gadiraju2023disability}. However, relatively few efforts have been made to systematically interrogate the power dynamics underlying these approaches, such as who or what perspectives determine the systems' goals, labeling frameworks, and what counts as the `correct' labels. As a result, the following ableist assumptions frequently surface in AI initiatives, even those intended to make AI more accessible:
\begin{enumerate}[itemsep=2pt, topsep=2pt]
    \item \textbf{Disability is often viewed as a medical deficit:} Many AI systems assume an ``ideal'' human standard and classify those deviate ``disabled'' and in need of fixing. For instance, extensive work relies on medical frameworks to categorize disability~\cite{song2019use, lea2021sep28k, cho2019review, schuller2013interspeech, mikolas2022training}, even though many diagnositic frameworks lack consensus or explanatory validity~\cite{apperly2024transdiagnostic, astle2022annual, yaruss1997clinical, beavan2011prevalence, pickersgill2014debating}. Medical frameworks also underpin systems designed to fit disabled bodyminds into able-bodied standards, such as correcting dysfluent speech~\cite{samsung-stuttering, dash2018speech} or training autistic people to exhibit neurotypical social behaviors~\cite{nagy2024autism, daniels2018exploratory, philip2025social}. Without critical examination and careful design, these systems may reinforce flawed medical models and perpetuate ableist worldviews.
    \item \textbf{Able-bodied people are often positioned as the authoritative source of knowledge}: Able-bodied people are often treated as the default reference for task goals or as the source of ground-truth labels. For instance, research on AI for blind and low-vision (BLV)\footnote{Blindness is a nuanced experience. We use blind and low-vision throughout to refer to visual disabilities broadly.} people has largely focused on translating visual information into non-visual modalities~\cite{gamage2023smart}, while paying less attention to non-visual forms of knowledge generated in BLV communities~\cite{heylighen2014designerly, hsueh2023visualization, protactile} such as tactile languages~\cite{clark-against-access}. Besides, these systems often rely on visual descriptions provided by sighted people for training~\cite{gurari2018vizwiz, gurari2020captioning}, while overlooking BLV people's sense-making strategies~\cite{tang2025everyday, garg2025s}.
    \item \textbf{Disabled people are often treated as passive objects of protection}: AI accessibility is typically treated as a matter of protection and accommodation, which casts disabled people primarily as sites for technological intervention rather than as co-creators~\cite{marathe2026rhetoric}. As a result, disabled people’s involvement is often limited to providing training data or post hoc feedback~\cite{haimson2025attitudes, phutane2025cold, glazko2024resume, gadiraju2023disability, mack2024representation}, with little influence over the core development decisions such as how their data should be collected, interpreted, and used.
\end{enumerate}
These limitations highlight the need for a more comprehensive framework to guide the development of AI for disability. Without careful consideration, the resulting AI models and systems may not only fall short in benefiting disabled communities but also reinforce ableism.

\vspace{-1em}
\subsection{Crip Theory and Cripping Tech}
Crip theory is a body of critical scholarship on disability emerging from feminist disability studies and queer theory~\cite{mcruer2008crip, kafer2013feminist}. Crip is a reclaimed slur, intentionally used as a political term to foreground the violence surrounding disability and to challenge normative assumptions about what bodyminds count as `normal' and what are not. Cripping uses ``crip'' as a verb to expose systems of normalcy and envision alternative ways of being that center disability. 

Crip theory, or critical disability studies more broadly, have been vital in shaping computing research such as HCI and accessible computing~\cite{janicki2024queering, jamshed2025productivity, janicki2024crip, riggs2025queer, homewood2025cripping, hsueh2023visualization, mankoff2010disability, williams2021articulations}. This line of work has revealed pervasive ableist assumptions embedded in technologies, and made efforts to transform design to center disability~\cite{hsueh2023visualization, janicki2024queering, jamshed2025productivity}. For example, Hsueh et al. challenged vision-centric norms in data visualization that `disable' BLV people, shifting the focus towards supporting BLV people in generating audio and tactile representations of data~\cite{hsueh2023visualization}. However, while the critical scholarship has informed the design of assistive technology more broadly~\cite{mankoff2010disability, bennett2018interdependence, williams2023counterventions}, how to truly center lived disability experiences in AI development remains exploratory.

\begin{table*}[h]
\centering
\footnotesize
\begin{tabularx}{\textwidth}{
    >{\raggedright\arraybackslash}p{2cm}
    >{\raggedright\arraybackslash}p{5cm}
    >{\raggedright\arraybackslash}X
}
\toprule
\textbf{Tenet} & 
\textbf{Key Ideas} &
\textbf{Implications for AI Research and Development} \\
\midrule

Expose disability politics & 
Interrogating how disability politics is encoded and reinforced in AI practices
\begin{itemize}[leftmargin=*, nosep, topsep=2pt]
    \item Instead of treating  ``disabled'' and ``non-disabled'' as fixed, recognizing disability as a socio-politically shaped experience emerging from situated environments (e.g., societal attitudes, architecture affordances).
    \item Challenging the assumption that there is one `ideal,' `normal' way humans should function.
    \item Recognizing disability as a valid human experience centering on disabled people's lived experiences.
\end{itemize}
&
\begin{itemize}[leftmargin=*, nosep, topsep=2pt]
    \item Contesting ableist norms and affirming disabled experiences (e.g., shifting from training autistic individuals to mimic neurotypical behaviors~\cite{voss2016superpower, daniels2018exploratory} toward supporting neurodivergent expressions~\cite{spiel2022adhd, williams2023counterventions, joseph2025redefining}).
    \item Moving from rigid diagnostic classifications (e.g., ~\cite{lea2021sep28k, mikolas2022training, chancellor2020methods}) toward a more fluid understanding of individual experiences (e.g., adopting transdiagnostic approaches to neurodiversity or mental health that center people's needs and experiences~\cite{astle2022annual, dalgleish2020transdiagnostic}).
    \item Questioning ableist assumptions in Artificial General Intelligence (AGI) narratives and practices (e.g., equalizing intelligence to fluency in English~\cite{smith2024standard}); and recognizing forms of intelligence generated through disabled bodyminds (e.g., non-verbal communication forms~\cite{henner2023unsettling, goodwin2004competent, alper2018inclusive}).
\end{itemize}
\\
\midrule

Honor cripistemologies & 
Centering disabled ways of knowing in AI development
\begin{itemize}[leftmargin=*, nosep, topsep=2pt]
    \item Challenging the idea that able-bodied people hold the authority on knowledge.
    \item Valuing knowledge produced through disabled bodyminds and communities.
\end{itemize}
&
\begin{itemize}[leftmargin=*, nosep, topsep=2pt]
    \item Reorienting assistive technology design from simulating able-bodied perceptions to supporting disabled ways of knowing (e.g., supporting blind users in triangulating information across different spatial perspectives such as allo- and ego-centric frameworks~\cite{tang2025everyday}).
    \item Incorporating disabled people's lived experiences and expertise to shape AI development~\cite{tang2026disability, tang2026reimagining, garg2025s} (e.g., instead of benchmarking sign language AI systems against hearing interpreters~\cite{desai2024systemic}, these systems should be built on lived deaf signing practices~\cite{tang2026reimagining}; allowing affected community members to interpret their own experiences and data~\cite{tang2026disability}).
    \item Grounding AI development and evaluations in disabled people's real-world practices, rather than relying solely on de-contextualized, isolated, time-bound benchmarks (e.g., instead of assuming model outputs as correct, blind people often situate their interpretation of model output in concrete tasks such as searching for a photo or reading a document~\cite{tang2025everyday}.)
\end{itemize}
\\
\midrule

Respect crip labor & 
Recognizing AI accessibility as an ongoing, friction-filled, collective work in which disabled people play a key role
\begin{itemize}[leftmargin=*, nosep, topsep=2pt]
    \item Treating accessibility as an access-making process, rather than a one-off technical solution.
    \item Recognizing the labor PWDs take in co-creating and sustaining accessibility, rather than treating them as passive recipients of assistance.
\end{itemize}
&
\begin{itemize}[leftmargin=*, nosep, topsep=2pt]
    \item Engaging community members to better understand their situated and evolving practices (e.g., BLV people often do not accept AI output as what it is but evolve their understanding of information within their workflows~\cite{tang2025everyday}).
    \item Designing accessibility as an ongoing, relational practice of care (e.g., ensuring human assistance when AI systems fail~\cite{moore2025executive}).
    \item Interrogating how AI systems shift labor and introduce new accessibility burdens for PWDs (e.g., personalized methods often requires extra effort from PWDs in data collection and model tuning~\cite{kacorri2017teachable, bragg2016sound, google-relate}).
\end{itemize}
\\
\bottomrule

\end{tabularx}
\caption{The framework we propose for cripping AI, including three interconnected tenets, the key ideas, and their implications for AI research and development.}
\vspace{-3em}
\label{table::framework}
\end{table*}

\vspace{-1em}
\section{Cripping AI}
\label{sec::cripping}
We propose three interconnected tenets as a starting point for systematically cripping AI: (1) exposing disability politics, (2) honoring cripistemologies, and (3) respecting crip labor. These tenets are offered as a generative framework rather than a prescriptive method. We provide an overview of our framework in Table \ref{table::framework} with more concrete case examples to illustrate how to practice these tenets in Section \ref{sec::cases}.

~\vspace{-1em}
\subsection{Expose disability politics: interrogating how power relations are encoded and reinforced in AI practices} 
Crip theory, including critical disability studies more broadly, centers on revealing and dismantling the power dynamics around disability, i.e., who or what perspectives get the power to define disability. This tradition rejects essentialist views of disability, arguing that disabled people are `disabled' through interactions with their surrounding environments -- ranging from medical systems to political structures (e.g., how disability is operationalized in census data~\cite{landes2024counting}), societal attitudes, and the material affordances of the built environment (e.g., ramps and elevators)~\cite{kafer2013feminist, padden1988deaf, linton2017reassigning, kim2017curative}. For example, diagnostic frameworks such as DSM are often deemed as the authoritative standard for understanding mental or neurodevelopmental disabilities~\cite{mikolas2022training, song2019use, cho2019review}. However, these standards have their limitations and have been repeatedly challenged by clinical professionals. Recent clinical work increasingly advocates for transdiagnostic approaches that shift the focus from diagnostic categories to individuals' needs and experiences~\cite{apperly2024transdiagnostic, astle2022annual}. The choice of standards therefore actively shapes who is considered disabled and who is not, while the lived experience of disability is further mediated by the interplay of environmental barriers and support available.

Furthermore, scholars in critical disability studies challenge assumptions about an `ideal' human standard and position disability as an intrinsic part of humanity~\cite{kafer2013feminist, mcruer2008crip, kim2017curative}. Robert McRuer introduced the term of \textit{compulsory ablebodiedness}~\cite{mcruer2008crip}, to highlight the socially enforced expectation that individuals should conform to normative ideals of being ``able-bodied'', or risk marginalization. Aligned with these perspectives, medical professionals increasingly emphasize the importance of including disabled people's perspectives in interpreting their own experiences and understanding the support they need~\cite{constantino2020speaker, joseph2025redefining, schrader2013mad}. A notable example is the neurodiversity movement, which challenges traditional biological and psychiatric standards of autism, arguing that it should be recognized as a form of neural diversity~\cite{singer2017neurodiversity}. Under this movement, there is growing recognition that behaviors often labeled as disorders or communication problems like repetitive behaviors, are autistic people's important communication methods~\cite{pinchevski2016autism, safira2020non}. The idea of neurodiversity has also expanded beyond autistic communities, gaining support among people who identify with ADHD, dyslexia, mental illness, etc.~\cite{mclennan2025neurodiversity}

\paragraph{Implications}
Following the crip tradition of interrogating disability politics, cripping AI should expose disability politics embedded in AI research and practices, and center disabled people's lived experiences instead. Take neurodivergence\footnote{We do not differentiate neurodivergent populations into diagnositic categories here as the validity of these traditional frameworks has been increasingly challenged~\cite{astle2022annual}.} as an example. Prior systematic reviews indicate that systems designed for neurodivergent individuals often emphasize diagnostic categorization and aim to align their behaviors with neurotypical norms, such as maintaining eye contact and regulating attention~\cite{spiel2019agency, williams2023counterventions, williams2020perseverations, philip2025social, spiel2022adhd}. Following these trends, recent work has applied generative AI (GenAI) models both for diagnosis and interventions~\cite{sohn2025implementation}. Research also indicates that large language models (LLMs), such as ChatGPT, encode stereotypes about neurodivergent people, such as assuming autistic people have social challenges~\cite{park2025paradox}. To avoid reinforcing the dominance of neurotypical norms, an important next step is to examine how neurodivergence is conceptualized such as in datasets~\cite{metcalf2025categorical} and evaluation measurements~\cite{giri2025do}, engage with neurodivergent communities to understand their perspectives~\cite{frauenberger2017blending, singer2017neurodiversity}, and transform existing practices to better support neurodivergent expression, from design goals to evaluation~\cite{spiel2019agency, williams2023counterventions}. For example, GenAI systems can support neurodivergent needs such as identification and management of sensory sensitivity~\cite{dotch2024accessibility}.

General-purpose AI models and the narratives about them may also perpetuate ableist assumptions. For example, critics argue that LLMs, like ChatGPT, prioritize ableist norms such as efficiency~\cite{jamshed2025productivity} or fluency in speech and writing~\cite{ciurria-ableism-gpt}. More broadly, the pursuit of Artificial General Intelligence (AGI) is often guided by a narrowed conception of intelligence and often evaluated using standardized benchmarks or by comparison to ``human baselines''~\cite{phan2025humanity, deepmind-baseline}. However, as critiques of human IQ tests have long argued~\cite{kamin2012science, madaio2022beyond, nature-intelligence}, these metrics and evaluation tasks may not truly measure ``general'' intelligence but instead privilege a limited set of social and cultural norms about what counts as intelligence and abilities (e.g., mathematical skills~\cite{phan2025humanity}, fluency in English~\cite{smith2024standard}). Failing to address these historical debts risks further marginalizing intelligence central to disability, for example, non-speech communication such as gaze, gesture, and embodied interaction~\cite{goodwin2004competent, henner2023unsettling, kusters2017beyond, alper2018inclusive}. Cripping AI therefore calls for deeper investigations into diverse forms of human intelligence to include those generated through disabled bodyminds.

\subsection{Honor cripistemologies: centering disabled ways of knowing in AI development}
Crip theorists also challenge the belief that able-bodied people have the ultimate authority on knowledge, a common assumption marginalizing the knowledge produced by disabled bodyminds and communities. Johnson and McRuer coined the term, \textit{cripistemology}, which combines \textit{crip} with \textit{epistemology} to highlight distinctive forms of knowledge generated through disability~\cite{johnson2014cripistemologies}. Cripistemologies can be found everywhere: in everyday innovations~\cite{wu2023good, oxo, buehler2015sharing, bennett2019biographical, hamraie2025crip, hamraie202329}, new technologies~\cite{gallaudet-tty, typewriter}, artistic styles, techniques, themes~\cite{kusama, kahlo, gruener2015effect, carelli2014painting, stevie-django-music, beethoven, signmark, kendall1988degas, molly-joyce}, and sports~\cite{yuru-sports, wheelchair-basketball, climbing}. Crucially, they do not stem from ``overcoming'' disability as how ableist frameworks interpret ``abilities,''~\cite{inspiration-porn, schalk2016reevaluating} but emerge naturally from engaging with the world through disabled embodiment. For example, Tony Iommi, the guitarist for Black Sabbath who lost two fingertips, adapted his guitar strings, modified his play techniques, and made fingertip covers. These adaptations not only facilitated his guitar playing but also produced a darker sound that many consider the origin of heavy metal~\cite{tony-iommi-pain}.

Besides creative hacking of spaces and tools, cripistemologies often emerge from disabled embodiment and are expressed through rich sensory and relational engagement with others and the environment. Critical disability theorists describe this idea through the concept of \textit{interdependence}~\cite{bennett2018interdependence, hsueh2023visualization}, which situates the access disabled people build in an expansive network of humans, objects, and environmental cues. For example, when a BLV person navigates a space, they do not rely on a single source of information, but actively gather environmental cues and shift in cognitive frameworks~\cite{saerberg2010just}, such as triangulating between ego- (e.g., ``to my right'') and allo-centric (e.g., ``east of the park'') spatial frameworks~\cite{giudice2018navigating}, interpreting tactile feedback from the ground~\cite{williams2014just}, using a white cane to detect environmental features~\cite{williams2014just}, and collecting interaction cues from a sighted guide if there is any~\cite{due2021distributed}. Ironically, what sighted people think is the right feedback is too often the wrong one for BLV people~\cite{williams2014just}. For example, many navigation systems are designed for obstacle avoidance and define the optimal path as the widest open space. However, objects on roads often serve as important informational cues for BLV people~\cite{williams2014just}, making the widest open space a worse environment for them.

\paragraph{Implications}
Drawing on the notion of cripistemologies, cripping AI challenges practices that treat able-bodied people as the primary source of knowledge for AI development~\cite{gurari2018vizwiz, gurari2020captioning, wu2017automatic, lea2021sep28k}. In the context of blindness, this tendency is evident in decades of work on having sighted crowdworkers describe visual content~\cite{wu2017automatic, yan2025chart} or translating visual information into non-visual modalities for BLV people~\cite{hamilton2016sensory}. Similarly, sign language systems are often modeled and benchmarked against interpreters (often hearing people who use sign language as a second language), while engaging minimally with deaf\footnote{Deaf communities in the U.S. often capitalize “D” in “Deaf” to emphasize cultural identity and pride, while using lowercase “deaf” to refer to hearing status. We use deaf throughout to acknowledge the fluidity of identity. We use deaf and hard-of-hearing when referring more broadly to people who have hearing disabilities.} signers~\cite{desai2024systemic}.

Moreover, many AI assistive applications are built on de-contextualized, isolated, and time-bound tasks. These traditional machine learning tasks often assume a static, single ground truth, for example, that each image has a single `accurate' description~\cite{gurari2018vizwiz}, each face expresses a specific type of emotion~\cite{daniels2018exploratory}, or that each English sentence has a direct parallel in sign language~\cite{desai2023asl}. While these efforts mark meaningful progress, relying on a single label can fall short in real-world scenarios, as the information needs of PWDs often vary across different contexts~\cite{stangl2021going, tang2026reimagining}. The single ground-truth paradigm might also mask the inherent subjectivity in human interpretation~\cite{tang2025everyday, tang2026reimagining}. Take visual description as an example. A vehicle expert may draw on a specialized vocabulary to describe a car, while a visual designer might include more detailed descriptions of colors~\cite{tang2025everyday}. As a result, BLV people must navigate incomplete and subjective information from both AI models and humans in their everyday information seeking, often developing their unique sense-making strategies~\cite{tang2025everyday}. 

Cripping AI therefore emphasizes grounding system design into disabled people's knowledge and real-world practices. For example, visual assistive models are frequently benchmarked against descriptions provided by sighted crowdworkers, who, however, often have no idea how BLV people understand visuals and how the image will ultimately be used~\cite{gurari2018vizwiz, gurari2020captioning}. To center blind knowledge in AI development, AI systems should lean on BLV users' knowledge frameworks to enhance their existing sense-making strategies. For example, a visual assistive application can provide descriptions from people with varied expertise to support BLV people in information triangulation~\cite{tang2025everyday}. A more radical approach would further shift the design focus away from vision-centered systems toward audio and tactile forms of knowledge. For example, instead of focusing on making data visualization more accessible to BLV people, systems could be built to support BLV people in creating audio and tactile representations of data (see~\cite{hsueh2023visualization, zong2024umwelt} for examples).

\vspace{-0.8em}
\subsection{Respect crip labor: recognizing AI accessibility as an ongoing, friction-filled, collective work}
\label{subsection::labor}
Furthermore, a crip perspective emphasizes the often-hidden labor and extra work disabled people perform to navigate an ableist world~\cite{price2024crip, zhang2026access}, including dealing with accommodations and the emotional labor involved~\cite{branham2015invisible, shinohara2021burden, zhang2026access}. This idea of crip labor challenges the assumption that disabled people are passive recipients of assistance and that accommodations inherently benefit them. Extensive evidence shows that creating accessibility is ongoing, friction-filled work in which disabled people actively participate~\cite{bennett2018interdependence, branham2015collaborative, bennett2020care, tang2025everyday, price2024crip}, often resulting in additional labor~\cite{branham2015invisible, shinohara2021burden} and without an ideal solution~\cite{tang2025everyday, price2024crip, tang2026designing}. A typical scenario is work and collaboration in mixed-ability teams, where members must deal with inaccessible tools and constantly negotiate between differing and sometimes conflicting needs~\cite{hofmann2020living, hsueh2023visualization, alharbi2023barriers, mack2021mixed, tang2026designing}. For instance, in video conferencing, BLV members might prioritize visual privacy, while deaf members might prefer to have visual cues; both need to navigate error-prone captions~\cite{alharbi2023barriers, mack2021mixed}. Disability activist Leah Lakshmi Piepzna-Samarasinha thus advocates for the principle of mutual aid to understand access practices~\cite{leah-collective}. The idea suggests that, instead of having a fixed solution, accessibility should be understood as an ongoing and sometimes messy practice to support one another.

\textit{Implications.}
Crip labor highlights accessibility as an access-making process with disabled people as key participants. This idea challenges the traditional view of accessibility as a one-off technical problem. Under this traditional view, the evaluation of AI systems might focuses on output accuracy or task completion~\cite{chang2024world, zhang2025towards, gurari2018vizwiz, gurari2020captioning}, and disabled people are usually limited to providing data or testing the final system~\cite{de-meulder-2021-good}. A typical example is the use of data on remote visual assistance platforms for training automated solutions. These platforms connect BLV users with sighted volunteers to answer visual questions~\cite{gurari2018vizwiz, gurari2020captioning}, while later becoming a major source of training AI systems for BLV people. While both human- and AI-based platforms benefit BLV communities, BLV users are often only involved in the one-time task requests, with little influence over other development stages. This approach misses the fact that BLV people's understanding of information often evolves in their interactions with human or tools. They constantly interpret and evaluate, not only AI outputs, but also the subjective nature of information (e.g., interpreting subjective questions like those about fashion, or the source of the information), while constantly dealing with inaccessible websites and artifacts~\cite{tang2025everyday, bigham2017effects}. Building visual assistive applications should therefore be a collaborative effort with BLV people to support their lived information practices. 

Framing access as a collaborative activity also points to the socio-emotional dimensions of access, such as trust and pleasure in co-creating accessible experiences~\cite{bennett2018interdependence}. For example, back-up channels like human assistance are necessary for people to receive continuous care and to build trust in a service when AI systems are limited~\cite{moore2025executive}. The trust involved in negotiating access also relies on a broader network of support, including policy frameworks that help sustain reliable access~\cite{knowles2023trustworthy}.

Furthermore, attending to crip labor calls for moving beyond surface-level accessibility claims and examining how AI systems shift labor and create new accessibility challenges. Although dominant narratives often frame AI as ``empowering'' disabled people, these systems can also introduce new accessibility issues~\cite{das2024from, adnin2024look, huh2024long}. For example, personalized methods have frequently been used to adapt general-purpose AI systems to disabled users' needs~\cite{theodorou2021disability, morrison2023understanding, kacorri2017teachable, google-relate, jain2022proto, bragg2016sound}. They typically require users to contribute data samples, such as recordings or images, to train or fine-tune the system~\cite{theodorou2021disability, morrison2023understanding, kacorri2017teachable, google-relate, jain2022proto}. While they improve AI performance, over-reliance on them risks shifting responsibility and labor onto disabled users themselves~\cite{tang2024privacy}, and diverting attention from more fundamental interventions. 

\vspace{-1em}
\section{Case Studies}
\label{sec::cases}
We present three cases from our own research to exemplify how to put our framework in practice. We begin with sign language AI and visual assistive AI as examples of systems targeting disabled communities, and turn to stuttering and speech AI as a case of general-purpose systems. For each case, we begin by introducing the relevant disability and systems, and then apply the three tenets.

~\vspace{-1em}
~\vspace{-1em}
\subsection{Deafness and Sign Language AI}
The first case draws on the first author's research on deaf communication and sign language technologies~\cite{tang2023community, tang2026reimagining}. Although deafness has traditionally been viewed as a `lack' of hearing, a cultural perspective views deaf people who use sign language as a linguistic and cultural minority~\cite{padden1988deaf}. A notable example is Martha's Vineyard, an island in the U.S. known for everyone there using sign language and non-signers appearing `disabled'~\cite{groce1985everyone}. From a cultural perspective, sign language AI is therefore a core technology to serve deaf communities. It typically encompasses techniques for sign language recognition, generation, and bidirectional translation with written or spoken language, often taking translation as the ultimate goal~\cite{bragg2019sign}. As research on responsible sign language AI~\cite{bragg2021fate, de-meulder-2021-good} and deaf-led initiatives~\cite{hickman2025bsl, desai2023asl, desai2024systemic} expand, cripping AI provides a framework to unify these efforts and surface remaining gaps in this area.

\textit{Tenet 1: Politics.} 
A crip perspective challenges hearing assumptions about language and communication and centers on deaf communication norms. Historically, hearing-centered beliefs have assumed spoken language as the only legitimate form of language, dismissed sign languages as a poor substitute, and pressured deaf children to use lip-reading and speech~\cite{dark-age}. It was not until the 1960s that sign languages were recognized as fully developed languages, when William Stokoe provided formal linguistic evidence demonstrating that ASL has its own distinct structures and vocabularies~\cite{stokoe2005sign}.

Despite increasing legal recognition of sign languages around the world~\cite{de2019legal}, technologies still often reduce their complexity due to lack of deaf awareness in design processes~\cite{glove_reject, tang2023community}. A notable example is sign language gloves, which are frequently promoted as devices capable of translating American Sign Language (ASL) into English characters through capturing handshapes~\cite{glove_reject, alex-lu-gloves}. While these technologies represented a significant step forward in sign language modeling, the idea faced strong push-back from deaf communities for missing crucial sign language elements like facial expressions~\cite{glove_reject, alex-lu-gloves}. In contrast to written/spoken languages that follow linear structures, sign language has its own unique, non-linear, spatial structures that incorporate manual (e.g., handshapes), non-manual elements (e.g., facial expressions), and bodily movements for expressions~\cite{glove_reject, alex-lu-gloves}. Recent research shows that misconceptions persist even among machine learning experts who have sign language processing experience ~\cite{kamikubo2025exploring}. This limited understanding may have led to widespread adoption of methods that oversimplify the language. For example, a recent review found that sign language systems often rely on annotation formats that reduce signs to simple tokens and do not follow best practices in linguistics~\cite{desai2024systemic}. The development of sign language AI therefore requires a shift in power to center signing communities' perspective and improve understanding of sign language linguistics. Hickman et al.'s work on British Sign Language offers a strong example of a deaf-led approach~\cite{hickman2025bsl}.

Developing sign language AI also involves understanding politics in the use and development of sign languages. As minority languages, sign languages often lack standardized forms and are shaped by complex histories and language ideologies~\cite{kusters2020sign, nakamura2006deaf}. Kusters et al. noted that if every regional and urban variety was given a distinct name, Indonesia alone would have over 500 named sign languages ~\cite{kusters2020sign}. Against these variants, deaf communities have heated debates over what counts as a `legitimate' sign language and how to sign `deaf' enough. A clear tension exists between sign languages naturally developed within deaf communities and manually coded languages (MCLs) created to represent spoken languages~\cite{schick2010development}. MCLs were not natural language but were artificial sign systems developed to help deaf children learn written/spoken languages. As a result, they are often criticized as ableist~\cite{nakamura2006deaf} and might even impact deaf children's language development~\cite{scott2019language}. For instance, Signing Exact English, represents English through letter-by-letter fingerspelling (consider sign `Starbucks' as `S-T-A-R-B-U-C-K-S'). Despite wide critiques, many signers may incorporate elements of these systems as these systems have already been used in deaf education~\cite{luetke1996history}. Similar to how Singaporeans blend English with other languages to develop Singlish~\cite{wei2018translanguaging}, the frequent language contact between sign language and written/spoken language systems then led to varieties such as Pidgin Signed English~\cite{hauser2000code}. These diverse and generative forms of sign languages challenge the idea that sign languages can be fully and inclusively represented in a standard model or dataset. Instead, a crip perspective shifts the focus to supporting deaf people's lived sign language practices, as we illustrate below.

\textit{Tenet 2: Cripistemologies.}
Deaf cripistemologies emerge from rich, community-developed signing practices~\cite{tang2023community, tang2026reimagining, friend-deaf}. As sign language information is often lacking in public spaces~\cite{tang2023community, padden1988deaf, tran2026toward}, deaf people have cultivated vibrant sign language spaces, e.g., deaf clubs~\cite{hickman2025bsl, padden1988deaf} and online communities~\cite{tang2026reimagining, tang2023community}. 

However, there remains a huge gap in incorporating deaf-led signing practices in sign language AI. Some of the long-standing benchmarks for sign language AI have been datasets curated from sign language interpreters~\cite{desai2024systemic}. However, the quality of sign language interpretation has been questioned in deaf communities, particularly when the interpreters are not actively engaged with deaf communities~\cite{de-meulder-2021-good, interpretation_tv, de2021sign, wehrmeyer2015comprehension, olson2017communication}. A survey of 10k+ deaf signers in China found that fewer than 10\% can fully understand sign language interpretation on TV broadcasting, as hearing interpreters usually follow Chinese grammar when signing (imagine someone translates ``quatre-vingts,'' 80 in French, as ``four-twenties'' for English speakers)~\cite{interpretation_tv}. The focus on interpreters in sign language AI research also raises questions: Who do these sign language technologies primarily serve? Are they aiming to serve deaf people, interpreters, or their hearing communication partners~\cite{de-meulder-2021-good}? Furthermore, the dominant focus on translation~\cite{desai2024systemic} may also limit the scope of signed language expressions to spoken and written languages~\cite{desai2024systemic}. To shift these power dynamics, future sign language systems should be grounded in translation practices within deaf communities~\cite{tang2023community, tang2026reimagining, friend-deaf}. Besides, more work should be done to support the flourishing of sign language itself, such as developing systems to support linguistics study~\cite{desai2024systemic}. 

The richness of sign language practices also questions the goal of sign language technologies to fully convert them into other languages. (Socio)linguists often draw on the term, \textit{translanguaging}, to describe deaf people's flexible use of languages~\cite{kusters2017beyond, henner2023unsettling}. The concept describes how bi- or multi-lingual speakers weave all their linguistic and non-linguistic resources to construct meaning~\cite{wei2018translanguaging}. Tang and Piper provided rich evidence showing how deaf content creators combine variants of sign languages, texts, captions, mouthing, images, and visual storytelling to translate information for diverse viewers~\cite{tang2026reimagining}. Sometimes singers even improvise and develop new signs when they encounter unfamiliar concepts or those from hearing cultures (e.g., brand names)~\cite{hin2022translanguaging, tang2026reimagining}. These practices are common in everyday signed communication, as signers frequently need to convey concepts from spoken languages that lack standardized signs~\cite{hin2022translanguaging, holmstrom2018deaf, tang2023community}, while also navigating complex internal linguistic variation~\cite{kusters2021emergence, tang2023community, tang2026reimagining}. These translanguaging practices not only challenge the idea of building translation machines for sign language, but also broaden the conventional scope of a named language to the whole semiotic space. Inspired by these deaf signing practices, sign language AI systems should similarly adopt modular architectures to support diverse languages, modalities, and cultural frameworks. The evaluation should focus on real communication scenarios and how well they fit existing deaf translanguaging practices.

\textit{Tenet 3: Crip Labor.}
Crip labor draws attention to deaf people's active roles in communication. There is rich evidence showing how deaf people co-construct and negotiate meaning with interpreters in communication, as well as with hearing or other deaf participants~\cite{wang2018accessibility, de2025deaf, clark-against-access, metzger1999sign}. For instance, Metzger demonstrated how interpretations unfold through collaboration, including jointly managing turn-taking and employing clarifications~\cite{metzger1999sign}. These negotiations challenge the idea that translation has a fixed ``ground-truth'' for benchmarking. To support communication as a collaborative activity, the development and evaluation of sign language AI should be grounded in deaf people's real-world communication scenarios (e.g., watching news, attending lectures), positioning AI systems as collaborative partners rather than mere language-conversion tools.

Negotiations in deaf communication, particularly when interpreters are involved, also challenge evaluation methods that depend solely on user feedback. As deaf scholar De Meulder noted, many deaf people have negotiated with low-quality interpretation services for their whole life, and they may see ``avatars' signing as another signing style they'll have to put up with and learn to `understand'''~\cite{de-meulder-2021-good}. Evaluating sign language AI therefore requires a team with diverse expertise including both lived experiences, linguistics, and translation or interpretation. More importantly, AI development should be grounded in the growth of sign language itself. For instance, future work can build on prior community efforts such as online signer communities~\cite{cavender2010asl, glasser2022asl, tang2023community} and deaf-led maker spaces~\cite{suchanek2025participation} to support collective development of sign language resources and translation practices.

Furthermore, crip labor problematizes the designs that cause extra work for deaf people. For instance, many sign language systems adopt wearables such as gloves or on-hand sensors to capture handshapes and movements~\cite{alex-lu-gloves, zhang2018finger}. However, these technologies may not only restrict signers' expressive flexibility but also reinforce the subordinate position of deaf people within hearing societies~\cite{alex-lu-gloves, gugenheimer2017impact}. To shift this dynamic, the development of sign language AI should prioritize systems that help hearing people learn to sign or directly benefit deaf communities. Examples include sign language wikis~\cite{glasser2022asl} and sign language dictionaries~\cite{bragg2015dictionary}.

~\vspace{-1em}
~\vspace{-1em}
\subsection{Blindness and Visual Assistive AI}
The second case draws on the first author's research experience on visual assistive AI for BLV people~\cite{tang2025everyday, garg2025s}. Similar to deafness, the definition and operationalization of blindness differ across legal, medical (e.g., visual acuity), and functional frameworks (e.g., what assistive technologies a person uses)~\cite{giudice2018navigating}. Despite these differences, visual assistive systems remain a primary focus in the development of AI for BLV people~\cite{gamage2023smart}. While varying in formats and use contexts, these systems share the goal to provide visual information to BLV people, for example, describing scenes~\cite{gamage2023smart}, generating image captions~\cite{gurari2020captioning}, answering visual questions about images~\cite{gurari2018vizwiz}, or identifying objects~\cite{kacorri2017people}. These systems typically use images provided by BLV individuals or sourced from the Internet for training, while relying on sighted crowdworkers to generate the corresponding visual descriptions. However, comparatively little attention has been given to the underlying assumptions in these efforts.

\vspace{-0.5em}
\paragraph{Tenet 1: Politics}
A crip perspective challenges vision-centered perspectives on blindness. For much of history, blindness has been defined in terms of a lack of sight, while vision has been positioned as the authoritative source of knowledge~\cite{schillmeier2006othering}. From this perspective, sighted crowdworkers were positioned as the sole source of `ground-truth' labels for visual assistive systems, often aiming to produce a single `accurate' and `unbiased' description~\cite{gurari2018vizwiz}. However, vision-centric frameworks not only overlook non-visual ways of knowing in support of visual-spatial understanding~\cite{heylighen2014designerly} but also oversimplify human vision~\cite{munton2022bias, goodwin2015professional, munton2022see, thierry2009unconscious}. Extensive evidence suggests that human vision is shaped by interconnected mechanisms of attention~\cite{ransom2017attention, carrasco2004attention}, memory~\cite{munton2022see}, professional training~\cite{goodwin2015professional}, language~\cite{thierry2009unconscious, winawer2007russian}, culture~\cite{masuda2001attending, segall1966influence, nisbett2005influence}, social structures~\cite{munton2019perceptual}, etc. Tang et al. describe how BLV people must navigate the inherent subjectivity in visual information from both AI systems and humans, a phenomenon they term ``everyday uncertainty'' ~\cite{tang2025everyday}. This pervasive subjectivity in visual perception questions the dominant paradigm that relies on sighted crowdworkers as the ground truth~\cite{gurari2018vizwiz}, especially considering that most sighted crowdworkers do not know what information BLV people need~\cite{simons2020hope}. As we elaborate below, rather than centering sighted knowledge, visual assistive AI should prioritize BLV people's information needs and recognize their multi-sensory strategies for accessing information.

\vspace{-0.5em}
\paragraph{Tenet 2: Cripistemologies}  
Blind cripistemologies manifest in rich non-visual sensory and interdependent ways of knowing, such as piecing together information from other people or tools to build understanding ~\cite{heylighen2014designerly, bennett2018interdependence}. Additionally, BLV people often have embodied knowledge unique to BLV communities, e.g., tactile features a Braille notetaker should have~\cite{johnson2025position}. However, these embodied forms of knowledge are often lacking in existing AI models and limit their abilities to represent blindness-related items and meet BLV people's needs. Research shows that both general-purpose and blind-focused AI systems often miss information crucial to BLV users (e.g., nutrition labels on products~\cite{garg2025s}) or assume sighted knowledge when generating visual descriptions~\cite{alharbi2025trying}. Additionally, many commonly used evaluation metrics of visual question answering systems are benchmarked against sighted preferences and do not consider BLV people's information needs~\cite{kapur-kreiss-2024-reference}. As an example of recent efforts to incorporate BLV perspectives in model development, Garg et al. conducted a survey to identify BLV users' information needs regarding products and re-annotated visual accessibility datasets to reflect these needs~\cite{garg2025s}. To advance fully blindness-centric AI models, future work should explore methods to support BLV communities in collaborative evaluation of visual descriptions~\cite{muehlbradt2022what}.

Existing visual AI assistive systems are also limited in supporting blind people's sense-making. These systems often assume a single, objective visual description that BLV users should receive. Although this assumption may work for some tasks (e.g., asking for a product brand when the product photo is clear and complete), it often fails in real-world situations that BLV users encounter, where the perceived utility of information varies according to the purpose of information seeking (curiosity vs. focused search), temporal constraints, and criticality (for healthcare vs. fun), etc.~\cite{tang2025everyday} The drive for an `accurate' AI output can even mask the interpretive nature of data, affecting BLV people's engagement with information. John Lee Clark, a Deafblind activist, illustrated this paradox with a story of working with a `biased' yet `best' interpreter~\cite{clark-against-access}. As he recalls, the interpreter was not professional and had strong personal opinions, but his descriptions were so direct and informative that Clark could interpret for himself what was happening in the room through the interpreter's perspective~\cite{clark-against-access}. By contrast, he critiqued the `accurate' alt-texts he typically encountered as ``a replica, divorced entirely from the original''~\cite{clark-against-access}. To avoid imposing a singular way of knowing, visual-assistive AI should focus on supporting BLV people's own sense-making strategies. For example, visual assistive AI systems should generate outputs from multiple sources and link other systems like human assistance and search engine to support BLV people in triangulation~\cite{tang2025everyday, hsueh2023visualization}. Furthermore, the evaluation should be grounded in their real-world workflows, such as how well they can support BLV users in searching for a photo or understanding a document~\cite{tang2025everyday}.

\vspace{-1em}
\paragraph{Tenet 3: Crip Labor}
From the lens of labor, while emerging AI tools offer potential support, they also introduce new forms of labor, such as managing erroneous outputs and prompt engineering~\cite{tang2025everyday}. The evaluation of visual assistive models should therefore prioritize reducing these forms of labor, such as focusing on errors BLV people care more about, such as recognizing restroom signs~\cite{ali2017embracing}. Besides, BLV users' use of AI models are often affected by embedded guardrails~\cite{tang2025everyday}. Companies set these guardrails in model outputs to prevent harms to user values, societal norms, and corporate interests~\cite{klyman2024acceptable}. However, these safeguards may also limit information that BLV users need, such as people's identities~\cite{hanley2021computer, bennett2021complicated} or potential risks in private spaces~\cite{tang2025everyday}. These frictions highlight accessibility as a collective work which needs a unified policy framework that accommodates accessibility with other values, such as privacy~\cite{tang2024privacy} and the interests of bystanders~\cite{akter2022shared}. 

Crip labor also draws attention to the power asymmetries in existing design approaches for BLV people. For instance, much research has focused on improving BLV people's photo-taking skills~\cite{sharma2023disability, kacorri2017people, morrison2023understanding, theodorou2021disability}, as they often take ``low-quality'' images that are considered hard for AI models to process, i.e., blurry, poorly framed, or missing key objects~\cite{gurari2018vizwiz, gurari2020captioning}. Yet, Garg et al. show that even imperfect photos often contain sufficient information for sighted people to identify valuable information such as product brands~\cite{garg2025s}. This raises fundamental questions: Who defines photo quality? Are these criteria aligned with the task, or do they merely reflect sighted assumptions? To shift dynamics, efforts should prioritize improving model performance for BLV people's information needs.

~\vspace{-1em}
~\vspace{-1em}
\subsection{Stuttering and Speech AI}
The third case draws on research conducted at AImpower.org (\url{https://aimpower.org/}), a nonprofit organization working with people who stutter (PWS). Our work involves understanding and improving PWS's experience with communication technologies such as videoconferencing~\cite{li2024reenvisioning, wu2023world, wu2024finding}, as well as developing disfluency-friendly speech AI through disfluent speech data collection and governance~\cite{li2024want, li2025collective, li2025govern, tang2026disability}, diagnostic model benchmarking~\cite{li2024towards, sridhar2025jjj}, and finetuning~\cite{li2025collective}. 
Stuttering is a neuro-developmental condition characterized by involuntary disruptions in speech, such as sound repetitions, word repetitions, sound prolongations, and blocks~\cite{stuttering}. Optimized for fluency and speed, speech AI systems including voice commands, speech-to-text tools, and automatic captioning often have great difficulties when interacting with PWS. 
%Stuttering represents an important but often neglected aspect of speech diversity in AI systems, such as voice commands, speech-to-text tools, and automatic captioning. These systems 
They often cut off or misunderstood stuttered speech, further marginalizing PWS in every aspects of daily life including employment and social interactions~\cite{lea2023from, sridhar2025jjj, li2024towards, wu2025speech}. To improve speech AI usability for PWS, previous efforts focused on improving automatic speech recognition~\cite{lea2023from, heeman2016using}, dysfluency detection~\cite{lea2021sep28k, kourkounakis2020detecting, sheikh2021stutternet}, and curating stuttered speech datasets~\cite{lea2021sep28k, batra2025boli, gong202470, li2024towards}. Despite these efforts, most of the work have focused on externally observable aspects of stuttering, with little attention given to the speaker's perspectives and experiences.

\paragraph{Tenet 1: Politics} 
A crip perspective challenges normative expectations of a ‘capable’ speaker, including fluency, immediacy, and efficiency~\cite{pierre2013construction}. These expectations create a hierarchy of fluent over disfluent speech, and have often led to speech systems built around fluency, including therapy programs that aim to correct disfluency~\cite{samsung-stuttering} and captioning systems that remove stuttered utterances by default~\cite{jiang-etal-2023-fluentspeech}. Recent research and advocacy in stuttering, however, have increasingly affirm  stuttering as a valid mode of communication~\cite{stammering-pride-2019}, defined not by speech disfluencies but by the speaker's subjective experiences, such as  spontaneity~\cite{constantino2020speaker} and embodied sensations~\cite{tichenor2019stuttering}. For example, PWS often emphasize the value of disfluencies as opportunities to foster empathy, trust, and deeper interpersonal connections~\cite{wu2023world, wu2024finding, constantino2020speaker}.  As Christopher Constantino (PWS, licensed SLP) observes, ``\textit{Our pauses, hesitations, and silences carry semantic weight; they are meaningful and purposeful}''~\cite{constantino-2016-stutter-gain}. Following these ideas, speech systems for PWS should shift goal from promoting fluency toward supporting the speaker's comfort and dignity in communication. For example, many PWS prefer having an option of verbatim captioning to normalize disfluencies as well as their identity as PWS~\cite{li2024reenvisioning, tang2026disability}.

Viewing stuttering as part of speech diversity also highlights the limits of relying solely on medical frameworks for speech modeling. Following the medical model, stuttered speech is typically evaluated by the frequency and duration of stuttering events~\cite{lea2021sep28k, sridhar2025jjj, gong202470, kourkounakis2021fluentnet}. The Stuttering Severity Instrument (SSI), one of the most popular clinical tools for measuring stuttering severity, requires speech-language pathologists (SLPs) to annotate the occurrence and a clear boundary of each stuttering event in stuttered speech samples~\cite{davidow2017intrajudge}.
%annotations of stuttered speech often attempt to define a clear boundary of stuttering. 
Similarly, computational modeling of stuttered speech has relied on labels of pre-defined stuttering event categories, such as blocks and repetitions, to represent stuttering~\cite{lea2021sep28k, sridhar2025jjj, gong202470, kourkounakis2021fluentnet}. However, stuttering measurement has been a contested issue even for SLPs due to its inter-rater variability and the overemphasis on observable speech behaviors~\cite{yaruss1997clinical}. For example, PWS often develop concealment behaviors to navigate moments of stuttering, such as interjections, word switching, or circumlocution~\cite{constantino2017rethinking}. These behaviors can blur the boundaries of stuttered and fluent speech, as a speaker can sound fluent by changing every word they block on and  a speaker can appear stutter as they pause intentionally. Consequently, annotating stuttering events has been found to be highly subjective, even for professionals~\cite{valente2025clinical}. To foreground PWS speakers' perspectives, our research shifts focus from clinical evaluations of their speech to real-world communication scenarios that matter to them, such as using voice commands and speaking in virtual meetings~\cite{li2024reenvisioning, tang2026disability}. For example, when we annotated stuttered speech datasets for ASR, we co-designed the annotation guidelines with PWS and invited PWS speakers to evaluate the quality of annotations on their own speech in concrete downstream applications~\cite{tang2026disability}. Our findings suggest that PWS speakers' expectations for annotations vary across communication contexts. For instance, some may care less about the accuracy of identifying specific stuttering events and more about whether auto captioning fosters awareness of stuttering. As a result, the ontology stuttering event labels is not fixed or strictly objective, but instead evolves through users' day-to-day use and interpretations.

Recognizing stuttering as a variable and dynamic communication experience helps surface the broader politics shaping how stuttered speech datasets are curated. Stuttering patterns fluctuate across communication partners, purposes, and environments~\cite{yaruss1997clinical}. Yet, most stuttered speech datasets are curated in lab settings, for example through tasks like reading passages aloud in front of researchers~\cite{ratner2018fluency, bayerl2022ksof, batra2025boli}. Such approaches often fail to capture how PWS speak across different contexts. Additionally, speaking in front of people without  lived experiences of stuttering may also make PWS feel less comfortable stuttering openly and authentically~\cite{lowe2021speech}. Building stuttering affirming AI therefore needs PWS' cripistemologies, as we elaborate below.

\vspace{-0.7em}
\paragraph{Tenet 2: Cripistemologies}
From a cripistemology perspective, PWS possess rich knowledge derived from their embodied experiences of stuttering. These forms of knowledge from our PWS team members and collaborators have proved essential to shaping our ongoing data work. For example, under their guidance, we chose unscripted conversations and dictation of voice commands as our collection tasks because they are two most frequent and critical scenarios in which PWS engage with speech technologies~\cite{li2025collective, li2024towards}. Collecting speech data in the form of unscripted conversations between PWS also grants the speakers full agency to engage in  topics that are meaningful to them, fostering narrative sovereignty for a population whose voices are often silenced in society~\cite{li2025collective}. PWS-led practices therefore allowed us to capture more versatile, naturalistic, and value-aligned stuttered speech data than datasets curated through expert-led paradigms~\cite{li2024towards, li2025collective}.

Our previous work also shows the importance of PWS's embodied knowledge to AI data annotation and interpretation. The iceberg theory of stuttering suggests that \cite{sheehan1970stuttering}, ``stuttering is like an iceberg, with only a small part above the waterline and a much bigger part below.'' Aligned with this theory, we found that PWS annotators disagreed with labels in more than 25\% of the clips in a highly cited English stuttered speech dataset~\cite{sridhar2025jjj, tang2026disability}. While non-PWS crowdworkers often follow rigid rules or focus on observable acoustic cues, PWS annotators can capture the bodily sensations that arise during and around moments of stuttering~\cite{tang2026disability}. For example, PWS annotators often drew on accessory and kinematic signals such as ``breathing, tone of speech, and speed of talking'' to inform their annotations~\cite{tang2026disability}. These nuanced perspectives from PWS challenge the accuracy of existing stuttered speech annotations produced by non-PWS, and also call into question some common practices in speech model training and evaluation, such as trimming segments classified as ``non-speech'' or ``silence'' (e.g., ~\cite{novitasari2022improving, zheng2025dncasr}), or using synthetic speech data to represent stuttering~\cite{kourkounakis2021fluentnet, zhang2025analysis}.

\vspace{-0.6em}
\paragraph{Tenet 3: Crip Labor}
Crip labor for PWS manifests across physical (e.g., bodily sensations), cognitive (e.g., negative thoughts), and emotional (e.g., anxiety) dimensions~\cite{constantino2020speaker}. These forms of labor uniquely shape how PWS participate in communication, which is often mediated by speech technologies. For example, the under-performance of ASR models could create additional labor for PWS such as fixing inaccurate transcriptions. %However, tolerance for errors differs depending on one's speech patterns: some people experience more repetitions, while others might primarily have blocks~\cite{tichenor2019stuttering}. 
However, the high variability within stuttering~\cite{tichenor2019stuttering} could lead to different types of model failures for different speakers: those who speak with more repetitions may see more hallucination in auto captioning~\cite{sridhar2025jjj}, while those with more blocks may experience frequent cut-offs by voice assistants~\cite{lea2023from}.
Fine-grained audits of speech AI models (e.g.~\cite{sridhar2025jjj, li2024towards, li2025collective} are thus required to ensure equitable performance within heterogeneous and intersectional groups such as the stuttering community.
%The evaluation of speech AI models should therefore be tailored to each PWS individual as performance on certain labels may impact some individuals more than others. 

PWS's labor should also be recognized in PWS's participation in AI development. For many PWS, confronting their own stuttering can trigger intense emotional responses~\cite{wu2023world}, imposing a significant emotional burden when they are asked to annotate or evaluate their own speech. These potential burdens highlight the importance of making dataset collection a rewarding experience for PWS data contributors. As an exploratory effort, our prior dataset work invited PWS community members or allies to take on interviewer roles and guide the data collection process~\cite{li2024want}. This PWS-centered approach not only helped foster an intimate and supportive environment for PWS participants, but also turned data collection into a rewarding community activity for personal growth~\cite{li2024want}. As a next step, we are investigating data stewardship frameworks to enable community members to oversee dataset curation and annotation~\cite{tang2026disability}.

\vspace{-1em}
\subsection{Summary of Insights}
\label{section::lessons}
Our cases offer three key insights to practice cripping AI as a guilding framework that centers lived disability experience across AI pipelines.
\begin{itemize}
    \item \textit{Lived-Experience as Expertise}: Lived disability experiences are critical expertise for AI development. In our three cases, lived disability experiences offer unique knowledge across language, information sense-making, and speech communication. These forms of embodied knowledge are crucial for moving beyond medical and able-bodied perspectives, building systems that truly support disabled people.
    \item \textit{Toward Lived-Experience-Preserving Pipelines}: Instead of assuming a single ground truth, AI systems should preserve the diversity, ambiguity, and subjectivity of human experience across data, training, and evaluation. Overall, this requires more reflexive and interpretive data practices (e.g., documenting different perspectives in dataset curation~\cite{newman2025disability, cambo2022positionality}), and system designs that support diverse viewpoints (e.g., allowing users to choose training data from different sources~\cite{gordon2022jury}, tailoring evaluations for individuals).
    \item \textit{Design with Lived-Experience in the Loop}: The full AI pipeline should co-evolve through ongoing collaboration with disabled communities. Our three cases show how disabled people interpret information in context, whether translation, visual information, or stuttering moments. These fluid, context-dependent practices require designing AI tasks and interpreting data within people's information practices and workflows, rather than relying on fixed training objectives.
\end{itemize}

\vspace{-1em}
\section{Future Directions}
\label{sec::discussion}
Cripping AI provides a framework to re-envision AI through positioning disability as a central dimension of human bodymind diversity. It not only strengthens existing calls for disability participation and accessible design but also demands genuine epistemic representation of knowledge produced through disabled bodyminds. Our framework suggests that collecting more data or soliciting feedback from PWDs is insufficient to make AI disability inclusive. AI development paradigms must expand to legitimize and value disabled experiences and knowledge. While the three cases illustrate how to crip AI in practice, we acknowledge that cripping AI should be an ongoing and collective effort. Below, we outline three directions to advance our framework.

\vspace{-1em}
\subsection{Toward a More Expansive and Plural Understanding of Human Bodyminds}
A direct next step is to develop and iterate our framework grounded in diverse lived disability experiences, including mental illness, chronic conditions, physical disabilities, and neurodivergence. Similar to the three cases we present, many disabled communities have been actively challenging traditional recovery models of disability and reclaiming their lived experiences as valid sources of knowledge for how they should be treated~\cite{rashed2019madness, chamberlin1978our, singer2017neurodiversity, miserandino-spoon-theory, kelly2022chronic}. These perspectives, however, have yet to be incorporated into AI development. For example, AI applications in mental health often operate within clinical frameworks, prioritizing the quantification and mitigation of ``symptoms'' like depression, anxiety, and distress~\cite{li2023systematic}. In contrast, mental health communities have cultivated Mad culture to promote patient-centered care and to frame fluctuations in emotions as part of their identity and natural human experiences~\cite{rashed2019madness, chamberlin1978our, kelly2022chronic}. Likewise, chronic illness and neurodivergent communities have affirmed fluctuations in energy as a valid dimension of human life~\cite{miserandino-spoon-theory, jennifer-spoon}. From these perspectives, AI tools should shift from assuming that human bodyminds should be stable toward supporting people in navigating and living with these fluctuations~\cite{ringland2019understanding, kornfield2020energy, jamshed2025productivity}. As the examples throughout this paper illustrate, cripping efforts should not be confined to specific communities or individuals. Reflecting on disabled experiences compels us to question our most fundamental assumptions about language, perception, communication, and well-being, and to transform AI practices to reflect the diversity of human bodyminds.

\vspace{-1em}
\subsection{Cripping the Entire AI Pipeline and Ecosystem}
Another critical direction is to transform the entire AI pipeline and ecosystem, from datasets, to computing architectures, benchmarks, evaluation measures, interfaces, etc. We offer three key recommendations in Section ~\ref{section::lessons} as a starting point. Given the scope limitation, we did not cover some crucial aspects such as model training, architecture, APIs, and governance. The increasing reliance on large foundational models makes these issues particularly challenging~\cite{suresh2024participation}, especially for accessibility applications with  limited data for training~\cite{desai2024systemic}. To address these challenges, AI systems must be designed with pluralism and multiplicity at their core, alongside robust auditing protocols as governance. Additionally, efforts should be made to crip broader digital infrastructures to ensure disabled communities' participation and oversight. For example, crowdwork platforms can create barriers for PWDs, such as requiring tasks to be completed within strict time limits~\cite{rechkemmer2022understanding}.

Ultimately, cripping the AI ecosystem requires shifting attention from individual users to the larger structural contexts in which AI systems operate. For instance, disabled individuals often face tensions between affirming their identities and meeting social norms~\cite{jang2024worker, marathe2025paradox, zhang2026access}. These tensions are likely to persist in the design of AI systems. For example, should AI systems provide advice to help autistic workers adapt to neurotypical workplace expectations~\cite{jang2024worker}? In these situations, a crip approach would argue for changing broader systematic ableist structures. For example, systems could be designed to help neurotypical colleagues better understand neurodivergent traits~\cite{haroon2025neurobridge}. A crip future therefore requires collaboration among all stakeholders, including but not limited company workers, company leadership, and healthcare professionals.

\vspace{-1em}
\subsection{Bridging Justice-Oriented AI Efforts}
Cripping AI also requires collaborating with other justice-oriented AI efforts including those driven by decolonialism~\cite{mohamed2020decolonial}, feminism~\cite{wellner2020feminist}, queer theory~\cite{queerinai}, critical race theory~\cite{ogbonnaya-Ogburu2020critical}, etc. This cross-community lens is essential for two reasons: (1) fostering a more careful, situated understanding of disability in practice, and (2) addressing the pervasive influence of ableism intertwined with broader intersectional power dynamics.

First, a cross-community approach is important as people experience overlapping systems of power~\cite{tang2025beyond}. For example, ASL is colonialized, racialized, and gendered~\cite{henner2023unsettling}. Deaf schools in the U.S. emerged from European settler colonial projects~\cite{henner2023unsettling}, which explains French Sign Language's influence over ASL. ASL was further influenced by racial segregation and gendered norms, leading to variations like Black ASL~\cite{hill2017importance} and gendered expressions (e.g., ``father'' is typically signed higher on the face than ``mother'')~\cite{asl-mom-dad}. A decolonial, anti-racist, crip sign language AI system should therefore avoid assuming there are idealized native speakers and standard ASL. Instead, it should vitalize deaf signers' living, embodied language practices under cross-community guidance.

Second, a cross-community approach is also essential because ableism not only affects disabled people but also sets narrow standards of ``ability'' that reinforces ideologies that disadvantage women, queer people, racial minorities, people in the Global South, etc. For instance, many AI hiring systems treat Western, masculine, and able-bodied traits as the default for professionalism, favoring candidates who maintain consistent eye contact, use expressive intonation, and speak fluent English with a ``standard'' accent~\cite{knight-job, rao2025invisible, bbc-job-hunting}. For training these systems, data workers are also faced with strict productivity standards, which often pressure them to squeeze diverse human behaviors into oversimplified labels~\cite{zhang2025making}. These structural issues reveal the intricate network of ableism, while highlighting the need for cross-community solidarity to drive systemic change.

%Ableism assumes idealized classifications of human bodies, reinforcing rigid categorizations in AI systems~\cite{scheuerman2021datasets}. ableism assumes productivity, devaluing labor like emotional...

\vspace{-1em}
\section{Concluding Thoughts}
This paper pictures a future in which AI is designed around the diversity of human bodyminds and ways of knowing. While we offer three cases to illustrate our approach, we acknowledge the risk of abstracting disabled experiences in our theorization and when we describe disabilities as a group. Additionally, we acknowledge that ableism remains a pervasive reality for disabled people~\cite{zhang2026access}, and that not all disabled individuals want to be called ``crip''~\cite{sherry-crip}. Claiming a crip identity and access to disability justice can itself reflect a form of privilege~\cite{sherry-crip, kirkham2021disability, mcdonnell2024understanding} and simplifies the histories and powers behind the word~\cite{shelley-crip, sherry-crip}. We therefore emphasize that our framework must remain situated and context-sensitive in practice. As our proposed future directions suggest, cripping AI must account for broad disabled communities and intersectional systems of power. Our framework seeks to serve as a starting point for this imperative.

%TC:ignore
\clearpage
\section*{Positionality Statement}
We acknowledge that our framework is deeply influenced by our identities and experiences. We are a group of researchers or practitioners based in U.S. academic or non-profit institutions. We have complex relations to multiple disabled identities and have academic training backgrounds in philosophy, psychology, human-computer interaction, accessibility, artificial intelligence, and critical disability studies. Our understanding of stuttering is informed by lived experience of the last author but none of us identify as deaf or blind. None of us are signers. Only the first author has basic knowledge of sign language. Our framework directly builds on our previous and ongoing research to understand lived disability experience and develop AI applications with various disabled communities. This work has spanned across diverse disabled populations in China, Europe, or U.S., including those who are deaf, BLV, neurodivergent, as well as those who have stuttering, and chronic illnesses. Our understanding of disability and AI is built upon years of work in this space, which spans across empirical studies~\cite{tang2025everyday, tang2026reimagining, li2024reenvisioning, zhang2026access, tang2023community}, AI data collection~\cite{li2024want, li2025collective}, annotation~\cite{tang2026disability}, governance~\cite{li2025govern}, model benchmarking and development~\cite{li2024towards, sridhar2025jjj, price2025lost, li2025collective}, as well as application design and development~\cite{li2024reenvisioning, wu2017automatic, wu2019design, zhao2017effect, zhao2018face}. 

\section*{Generative AI Usage Statement}
We used OpenAI ChatGPT, Google Gemini, and Perplexity to assist with grammar and style editing and information search (e.g., searching for examples of cripistemologies or references). All outputs were carefully reviewed and verified by the authors.

\begin{acks}
We are grateful to our colleagues and collaborators for their contributions to our earlier research. We thank Adryana Hutchinson for sharing the example of Tony Iommi as a great case of cripistemology. We also thank the reviewers for their constructive feedback on our earlier drafts. This work was supported by NSF Award \#2427710.
\end{acks}

\bibliographystyle{ACM-Reference-Format}
\bibliography{main}
%TC:endignore
\end{document}